\documentclass[12pt]{article}
\usepackage[utf8]{inputenc}
\usepackage[english, russian]{babel}
\usepackage{amsmath}
\usepackage{amsfonts}
\usepackage{amssymb}

\def\lb{\label}
\def\be{\begin{equation}}
\def\ee{\end{equation}}
\def\qed{\rule{5pt}{5pt}}

\newtheorem{proposition}{Proposition}

\begin{document}

\vspace{3cm}

	\begin{center}
		{\LARGE {Polarization spin-tensors in two-spinor formalism 

\vspace{0.2cm}

and Behrends-Fronsdal spin projection operator 

\vspace{0.3cm}

for $D$-dimensional case.}}

 \vspace{1cm}

\large \sf
M.A.~Podoinitsyn \footnote{\sf e-mail: mikhailpodoinicin@gmail.com} \\

\vspace{1cm}

{\it Bogoliubov Laboratory of Theoretical Physics, JINR, Dubna 141980, Russia}
\end{center}

\vspace{2cm}
\begin{otherlanguage}{english}
\begin{abstract}
\noindent
In the work, the recurrent differential relations that connecting the polarization spin-tensor 
of the wave function of a free massive particle of an arbitrary spin for $D=4$
and new formula of the $D$-dimensional Behrends-Fronsdal spin projection operator are found.
\end{abstract}
\end{otherlanguage}

\vspace{3cm}

\newpage

\section{Introduction}
\setcounter{equation}{0}
This work is a continuation of the article \cite{IP}. 
The  \cite{IP} is devoted to a two-spinor description 
of free massive particles of arbitrary 
spin and to the Berends-Fronsdal projection 
operators -- a projector onto irreducible 
completely symmetric representations of the $D$-dimensional Poincaré group. 
Each of these two sections received a small addition in the present work. 

 We briefly recall the main results of the \cite{IP}.
 We use the Wigner unitary representations
 of the group $ISL(2,C)$, which covers the Poincaré group.
 These representations are irreducible and one can reformulate
 them in such a way that these irreps act
 in the space of spin-tensor wave functions $\psi^{_{(r)}}$ of a special type.
 The construction of the functions $\psi^{_{(r)}}$
 is carried out with the help of Wigner operators, 
 which translate the unitary massive
 representation of the group  $ISL(2,C)$
 (induced from the irreducible representation of the stability
 subgroup $SU(2)$) acting in the space of Wigner wave functions $\phi$
 to a representation of the group  $ISL(2,C)$, acting in the space of
 special spin-tensor fields $\psi^{_{(r)}}$ of massive particles.
 In addition, the generalization on arbitrary dimension $D$ of the four-dimensional 
 Behrends-Fronsdal spin projection operator was found.

In the first section of this article, for fixing the notation and material consistency, 
we present the definitions of the spin-tensor wave function $\psi^{_{(r)}}$
the polarization spin-tensors $\overset{_{(m)}}{e}(k)$
and the expansion formula for $\psi^{_{(r)}}$ to
the sum over the polarization spin-tensors.
Further, using a special parametrization of Wigner operators in terms of 
two Weyl spinors, we prove the main Proposition {\bf \ref{Smr} } of the first section.
On the existence of differential recurrence relations connecting 
various polarization spin-tensors $j$.
These relations allow us to write out explicit expressions 
for the polarization spin-tensors $\overset{_{(m)}}{e}(k)$
in terms of two spinors. 

In the second part of the paper, we describe a new method 
for constructing of the Berends-Fronsdal spin projection operator 
for $D$-dimensional case.

\section{Polarization spin-tensors for the field of arbitrary spin.\label{VectPol}}

\setcounter{equation}{0}
Based on the Wigner  construction of massive unitary and irreducible
representations of the covering Poincare group $ISL(2,\mathbb {C})$,
one can show (see \cite{IP}) that the space of the unitary representation of the
group $ISL(2,\mathbb{C})$ with spin $j$ is transformed
to the space of the spin-tensor wave
functions of $ (\frac{p}{2},\frac{r}{2})$ - type depending on four-momentum $k=(k_0,k_1,k_2,k_3)$:
\begin{equation}
 \label{tpn}
\psi^{_{(r)}(\dot{\beta_1}...\dot{\beta_r})}
_{(\alpha_1...\alpha_p)}(k)=\frac{1}{{\sf m}^r}
 \prod_{i=1}^p (A_{(k)})^{\;\; \delta_i}_{\alpha_i}
\cdot  \Bigl( \prod_{j=1}^r
 \bigl(A^{-1\dagger}_{(k)}\bigr)^{\dot{\beta}_{j}}_{\;\; \dot{\xi}_j}
 \cdot (q^n \tilde{\sigma}_n)^{ 
 \dot{\xi}_j \delta _{p+j}} \Bigr) \;
 \phi_{(\delta _1...\delta _p \delta_{p+1} ...\delta_{p+r} )}(k) \; .
\end{equation}
Here $(p+r)=2j$, $\phi_{(\delta_1...\delta_p \delta_{p+1}...\delta_{p+r})}(k)$
-- is an arbitrary symmetric tensor of rank $2j$ (the Wigner wave function),
 $q^n$ -- are components of the test four-momentum $q = (q^0,q^1,q^2,q^3)$:
$$
q^n q_n= q_k \, \eta^{kn} \, q_n= q_0^2 - q_1^2-q_2^2-q_3^2 = {\sf m}^2 \; ,
\;\;\; \eta^{kn} = {\rm diag}(+1,-1,-1,-1) \; ,
$$
parameter ${\sf m} >0$ is the mass, $\tilde{\sigma}_n=(\sigma_0,-\sigma_1,-\sigma_2,-\sigma_3)$ and
 $\sigma_1,\sigma_2,\sigma_3$ -- are Pauli matrices while
 $\sigma_0$ -- is the unit $(2\times 2)$ matrix.
Matrices $A_{(k)},\; A_{(k)}^{\dagger -1}\in SL(2,\mathbb{C})$,
used in (\ref{tpn}), are solutions of the equations
\begin{equation}\label{ssm0n}
(A_{(k)})_\alpha^\gamma(q^n\sigma_n)_{\gamma \dot{\alpha}}(A_{(k)}^{\dagger})^{\dot{\alpha}}_{\dot{\gamma}}=
(k^n\sigma_n)_{\alpha \dot{\gamma}} \; .
\end{equation}
The matrix $A_{(k)}$ parametrizes coset space $SL(2,\mathbb{C})/SU(2)$.
The upper index $(r)$ of the spin-tensors $\psi^{_{(r)}}$ in (\ref{tpn})
 distinguishes these spin-tensors with respect to the number of dotted indices.
 In eq. (\ref{tpn}) we use operators $A_{(k)}^{\otimes p} \otimes
 \bigl(A^{\dagger -1}_{(k)}(q \tilde{\sigma})
 \bigr)^{\otimes r}$ to translate the Wigner wave functions $\phi$
  into spin-tensor functions $\psi^{_{(r)}}$ of
 $(\frac{p}{2},\frac{r}{2})$-type, 
 These operators are called {\it the Wigner operators}.

 According to \cite{IP}, the spin-tensor wave functions $\psi^{_ {(r)}} (k)$ 
 can be represented as the following sum over the polarization spin-tensors $\overset {_ {(m)}}{e} (k)$.
\begin{equation}\label{lk1}
\psi^{_{(r)}(\dot{\beta_1}...\dot{\beta_r})}
_{(\alpha_1...\alpha_p)}(k)=\frac{1}{\sqrt{(2j)!}} \;
 \sum_{m=-j}^j \phi_m(k) \,
\overset{_{(m)}}{e}^{(\dot{\beta_1}...\dot{\beta_r})}_{(\alpha_1...\alpha_p)}(k)
\; .
\end{equation}
The explicit form of the coefficients $\phi_m (k)$ is not needed 
here (you can see it in \cite{IP}), if the test momentum $q$ 
is fixed as $q = ({\sf m}, 0, 0, 0) $ polarizations $ \overset {_ {(m)}}{e} (k)$ are given by
\begin{equation}\label{npkp}
\overset{_{(m)}}{e}^{(\dot{\beta_1}...\dot{\beta_r})}_{(\alpha_1...\alpha_p)}(k)=
\frac{1}{\sqrt{(2j)}!}\prod^p_{i=1}
(A_{(k)})^{\;\;\; \rho_i}_{\alpha_i}\,
\prod^r_{\ell=1}
\bigl(A^{-1\dagger}_{(k)}\; \tilde{\sigma}_0\bigr)^{\dot{\beta}_{\ell} \rho_{p+\ell}}\,
\epsilon^{(m)}_{\rho_1\cdots\rho_{2j}},
\end{equation}
where we introduced
\begin{equation}\label{lk3}
\epsilon^{(m)}_{\rho_1\cdots\rho_{2j}}=\partial^{_{(v)}}_{\rho_1} \cdots \partial^{_{(v)}}_{\rho_{2j}} \; T_m^j(v) \; ; \;\;\;
T_m^j(v)=\frac{(v^1)^{j+m}(v^2)^{j-m}}{\sqrt{(j+m)!(j-m)!}} \; , \;\;\;
\partial^{_{(v)}}_{\rho}=\frac{\partial}{\partial v^{\rho}} \; ,
\end{equation}
where  $v^1, \; v^2$ the componets of the auxiliary Weyl Spinor.
\begin{proposition}\label{Smr}
The spin-tensors $\overset{_{(m)}}{e}^{(\dot{\beta_1}...\dot{\beta_r})}_{(\alpha_1...\alpha_p)}$, defined in (\ref{npkp})
satisfy the relations:

\begin{equation} \label{Smr31}
\bigl (\mu_\gamma \frac{\partial}{\partial \lambda_\gamma} -
\overline{\lambda}^{\dot{\gamma}} \frac{\partial}{\partial \overline{\mu}^{\dot{\gamma}}}\bigr)
\overset{_{(m)}}{e}^{(\dot{\beta_1}...\dot{\beta_r})}_{(\alpha_1...\alpha_p)} =
\sqrt{(j-m)(j+m+1)} \cdot \overset{_{(m+1)}}{e}^{(\dot{\beta_1}...\dot{\beta_r})}_{(\alpha_1...\alpha_p)}
\end{equation}

\begin{equation}\label{Smr3}
\bigl (\lambda_\gamma \frac{\partial}{\partial \mu_\gamma} -
\overline{\mu}^{\dot{\gamma}} \frac{\partial}{\partial \overline{\lambda}^{\dot{\gamma}}}\bigr)
\overset{_{(m)}}{e}^{(\dot{\beta_1}...\dot{\beta_r})}_{(\alpha_1...\alpha_p)}=
\sqrt{(j+m)(j-m+1)} \cdot \overset{_{(m-1)}}{e}^{(\dot{\beta_1}...\dot{\beta_r})}_{(\alpha_1...\alpha_p)}
\end{equation}

\begin{equation} \label{Smr32}
\frac{1}{2} \bigl (\mu_\gamma \frac{\partial}{\partial \mu_\gamma} -
\lambda_\gamma \frac{\partial}{\partial \lambda_\gamma}+
\overline{\lambda}^{\dot{\gamma}} \frac{\partial}{\partial \overline{\lambda}^{\dot{\gamma}}}
-\overline{\mu}^{\dot{\gamma}} \frac{\partial}{\partial \overline{\mu}^{\dot{\gamma}}}\bigr)
\overset{_{(m)}}{e}^{(\dot{\beta_1}...\dot{\beta_r})}_{(\alpha_1...\alpha_p)}=
m \cdot \overset{_{(m)}}{e}^{(\dot{\beta_1}...\dot{\beta_r})}_{(\alpha_1...\alpha_p)}
\end{equation}
where $\mu_{\gamma}, \, \lambda_{\gamma}, \,  \overline{\mu}^{\dot{\gamma}}, \,   \overline{\lambda}^{\dot{\gamma}}$ - Weyl spinors.
\end{proposition}
{\bf Proof.} The proof is based on the use the representation of matrices
$A_{(k)}, \, A_{(k)}^{-1 \dagger}\in SL(2,\mathbb{C})$ in terms of Weyl spinors $\mu, \lambda$
\begin{equation}
 \label{aonC}
 \begin{array}{c}
(A_{(k)})_{\alpha}^{\;\; \beta}=
 \frac{1}{z}
\begin{pmatrix} \mu_1 & \lambda_1 \\ \mu_2 & \lambda_2 \end{pmatrix} \; , \;
(A^{\dagger -1}_{(k)})^{\dot{\alpha}}_{\;\; \dot{\beta}} =
\frac{1}{z^*}
\begin{pmatrix} \overline{\lambda}_{\dot{2}} & -\overline{\mu}_{\dot{2}} \\ -\overline{\lambda}_{\dot{1}} & \overline{\mu}_{\dot{1}} \end{pmatrix}
 , \;\;\; \\ [0.4cm]
(z)^2 = \mu^\rho \, \lambda_\rho \, , \;\; (z^{*})^{2} =
\overline{\mu}^{\dot{\rho}}\, \overline{\lambda}_{\dot{\rho}} \, ,
\end{array}
\end{equation}
\begin{equation}
 \label{aonC1}
\mu=
\begin{pmatrix}
\mu_1 \\ \mu_2
\end{pmatrix} \; ,
\;
\lambda=
\begin{pmatrix}
\lambda_1 \\ \lambda_2
\end{pmatrix} \; ,
\;
\overline{\mu} =
\begin{pmatrix}
\overline{\mu}_{\dot{1}} \\ \overline{\mu}_{\dot{2}}
\end{pmatrix} \; ,
\;
\overline{\lambda} =
\begin{pmatrix}
\overline{\lambda}_{\dot{1}} \\ \overline{\lambda}_{\dot{2}}
\end{pmatrix} \, , \;\;\;\;
\overline{\lambda}_{\dot{\alpha}}=(\lambda_\alpha)^* \, , \;\;\;\;
\overline{\mu}_{\dot{\alpha}}=(\mu_\alpha)^* \; .
\end{equation}
proposed in paper \cite{IP}.
Let us show, for example, how one can prove of relation (\ref{Smr3}).
The proofs of relations (\ref{Smr31}), (\ref{Smr32}) are similar.
First of all we consider the obvious identity which follows
 from definitions (\ref{lk3}):
\be \label{Cheq1}
\epsilon^{(m-1)}_{\rho_1\cdots\rho_{2j}} =
\frac{\partial^{_{(v)}}_{\rho_1} \cdots \partial^{_{(v)}}_{\rho_{2j}} \Bigl(
 v^2 \, \partial^{_{(v)}}_{1} \; T_m^j(v)\Bigr)}{\sqrt{(j+m)(j-m+1)}}.
\ee
Now we expand the numerator of the right-hand side of the formula (\ref{Cheq1})
\be \label{Cheq2}
\begin{array}{c}
\partial^{_{(v)}}_{\rho_1} \cdots \partial^{_{(v)}}_{\rho_{2j}} \Bigl(v^2 \partial^{_{(v)}}_1 T_m^j(v)\Bigr) =
\Bigl( \delta^2_{\rho_{2j}} \partial^{_{(v)}}_{\rho_1} \cdots \partial^{_{(v)}}_{\rho_{2j-1}}+\cdots
+\delta^2_{\rho_{k}} \partial^{_{(v)}}_{\rho_1} \cdots \partial^{_{(v)}}_{\rho_{k-1}} \partial^{_{(v)}}_{\rho_{k+1}}\cdots \partial^{_{(v)}}_{\rho_{2j}}
\\[0.5cm]
+\delta^2_{\rho_{1}} \partial^{_{(v)}}_{\rho_2} \cdots \partial^{_{(v)}}_{\rho_{2j}}\Bigr) \partial^{_{(v)}}_1 T_m^j(v) =
\Bigl({\displaystyle \sum^{2j}_{d=1}  \delta^2_{\rho_{d}} \partial^{_{(v)}}_1 \prod^{2j}_{\underset{(n \neq d)}{n=1}}} \partial^{_{(v)}}_{\rho_n}\Bigr) \, T_m^j(v)
\end{array}
\ee
Then we substitute (\ref{Cheq1}) into (\ref{npkp}) and use (\ref{Cheq2})
\be \label{Cheq3}
\begin{array}{c}
 \overset{_{(m-1)}}{e}^{(\dot{\beta_1}...\dot{\beta_r})}_{(\alpha_1...\alpha_p)}(k)=\frac{1}{\sqrt{(j+m)(j-m+1)}}
\frac{1}{\sqrt{(2j)}!}
 {\displaystyle \prod^p_{i=1}
(A_{(k)})^{\;\;\; \rho_i}_{\alpha_i}\,
\prod^r_{\ell=1}
\bigl(A^{-1\dagger}_{(k)}\; \tilde{\sigma}_0\bigr)
^{\dot{\beta}_{\ell} \rho_{p+\ell}} \cdot} \\
{\displaystyle \Bigl( \sum^{p}_{d=1}  \delta^2_{\rho_{d}} \partial^{_{(v)}}_1
\prod^{p}_{\underset{(n \neq d)}{n=1}} \partial^{_{(v)}}_{\rho_n}
\prod^{2j}_{l=p+1} \partial^{_{(v)}}_{\rho_l}+
\sum^{2j}_{d=p+1}  \delta^2_{\rho_{d}} \partial^{_{(v)}}_1 \prod^{p}_{n=1}  \partial^{_{(v)}}_{\rho_n}
\prod^{2j}_{\underset{(l \neq d)}{l=p+1}} \partial^{_{(v)}}_{\rho_l}
\Bigr) \, T_m^j(v),
}
\end{array}
\ee
here we have divided the sum over $d$  in two parts (it will be needed for further consideration).
We also need the following identities
\be
 \lb{psm05}
\begin{array}{c}
 \mu_\alpha = z \, (A_{(k)})^{\;\;\; 1}_\alpha \, , \;\;
 \lambda_\alpha = z \, (A_{(k)})^{\;\;\; 2}_\alpha \, , \;\;
\\[0.2cm]
 \overline{\mu}{}^{\dot{\alpha}} = z^* \,
 (A_{(k)}^{\dagger -1}\; \tilde{\sigma}_0)^{\dot{\alpha} \;2} \, , \;\;
 \overline{\lambda}{}^{\dot{\alpha}} = - z^* \,
 (A_{(k)}^{\dagger -1}\; \tilde{\sigma}_0)^{\dot{\alpha} \;1} \, ,
\end{array}
 \ee
which follow from (\ref{aonC}) and (\ref{aonC1}). Note that the relations are holds
\be \label{zdc1}
\lambda_\gamma \frac{\partial z}{\partial \mu_\gamma} = \overline{\mu}_{\dot{\gamma}} \frac{\partial z^{*}}{\partial \overline{\lambda}_{\dot{\gamma}}}  = 0.
\ee
From the identities (\ref{psm05}) and the relations (\ref{zdc1}) we can get formulas
\be \label{zdc2}
\frac{\lambda_{\alpha}}{z} = \lambda_{\gamma} \frac{\partial}{\partial \mu_{\gamma}} (A_{(k)})^{\;\;\; 1}_{\alpha}, \;\;\;
\frac{\overline{\mu}^{\dot{\beta}}}{z^*} =-\overline{\mu}^{\dot{\gamma}} \frac{\partial}{\partial \overline{\lambda}^{\dot{\gamma}}}
\bigl(A^{-1\dagger}_{(k)}\; \tilde{\sigma}_0\bigr)^{\dot{\beta} \; 1}. 
\ee
Using first identities (\ref{psm05}) and then (\ref{zdc2}), the right-hand side of formula (\ref{Cheq3}) (without numeric factor and  
monomial $T_m^j$) can be rewritten as follows

\be \label{Cheq4}
\begin{array}{c}
{\displaystyle \biggl( \sum^{p}_{d=1}  \Bigl (  \lambda_{\gamma} \frac{\partial}{\partial \mu_{\gamma}} (A_{(k)})^{\;\;\; 1}_{\alpha_d} \Bigr ) \,
 \partial^{_{(v)}}_{1} 
\prod^{p}_{\underset{(n \neq d)}{n=1}}
(A_{(k)})^{\;\;\; \rho_n}_{\alpha_n} \,
\partial^{_{(v)}}_{\rho_n}
\prod^{2j}_{\ell=p+1}
\bigl(A^{-1\dagger}_{(k)}\; \tilde{\sigma}_0\bigr)^{\dot{\beta}_{\ell-p} \rho_{\ell}}
\partial^{_{(v)}}_{\rho_{\ell}} +} \\[0.5cm]
{\displaystyle \sum^{2j}_{d=p+1} \Bigl (   \overline{\mu}^{\dot{\gamma}} \frac{\partial}{\partial \overline{\lambda}^{\dot{\gamma}}}
\bigl(A^{-1\dagger}_{(k)}\; \tilde{\sigma}_0\bigr)^{\dot{\beta}_{d-p} \; 1} \Bigr ) \,
\partial^{_{(v)}}_{1} 
\prod^{p}_{n=1} (A_{(k)})^{\;\;\; \rho_n}_{\alpha_n}  \partial^{_{(v)}}_{\rho_n}
\prod^{2j}_{\underset{(\ell \neq d)}{\ell=p+1}}
 \bigl(A^{-1\dagger}_{(k)}\; \tilde{\sigma}_0\bigr)^{\dot{\beta}_{\ell-p} \rho_{\ell}}
 \partial^{_{(v)}}_{\rho_{\ell}}
\biggr) \, .
}
\end{array}
\ee
Now we add to (\ref{Cheq4}) the following zero terms
\be \label{Cheq5}
\begin{array}{c}
{\displaystyle \biggl ( \sum^{p}_{d=1}   \Bigl(  \lambda_{\gamma} \frac{\partial}{\partial \mu_{\gamma}} (A_{(k)})^{\;\;\; 2}_{\alpha_d} \Bigr ) \,
 \partial^{_{(v)}}_{2} 
\prod^{p}_{\underset{(n \neq d)}{n=1}}
(A_{(k)})^{\;\;\; \rho_n}_{\alpha_n} \,
\partial^{_{(v)}}_{\rho_n}
\prod^{2j}_{\ell=p+1}
\bigl(A^{-1\dagger}_{(k)}\; \tilde{\sigma}_0\bigr)^{\dot{\beta}_{\ell-p} \rho_{\ell}}
\partial^{_{(v)}}_{\rho_{\ell}} +} \\[0.5cm]
{\displaystyle \sum^{2j}_{d=p+1} \Bigl (   \overline{\mu}^{\dot{\gamma}} \frac{\partial}{\partial \overline{\lambda}^{\dot{\gamma}}}
\bigl(A^{-1\dagger}_{(k)}\; \tilde{\sigma}_0\bigr)^{\dot{\beta}_{d-p} \; 2}  \Bigr ) \,
\partial^{_{(v)}}_{2}
\prod^{p}_{n=1} (A_{(k)})^{\;\;\; \rho_n}_{\alpha_n}  \partial^{_{(v)}}_{\rho_n}
\prod^{2j}_{\underset{(\ell \neq d)}{\ell=p+1}}
 \bigl(A^{-1\dagger}_{(k)}\; \tilde{\sigma}_0\bigr)^{\dot{\beta}_{\ell-p} \rho_{\ell}}
 \partial^{_{(v)}}_{\rho_{\ell}}
\biggr) \, .
}
\end{array}
\ee
And as a result the sum of (\ref{Cheq4}) and (\ref{Cheq5}) has the form
\be \label{Cheq6}
\begin{array}{c}
{\displaystyle \biggl ( \sum^{p}_{d=1}\Bigl ( \lambda_{\gamma} \frac{\partial}{\partial \mu_{\gamma}} (A_{(k)})^{\;\;\; \rho_d}_{\alpha_d} \Bigr ) \,
 \partial^{_{(v)}}_{\rho_d} 
\prod^{p}_{\underset{(n \neq d)}{n=1}}
(A_{(k)})^{\;\;\; \rho_n}_{\alpha_n} \,
\partial^{_{(v)}}_{\rho_n}
\prod^{2j}_{\ell=p+1}
\bigl(A^{-1\dagger}_{(k)}\; \tilde{\sigma}_0\bigr)^{\dot{\beta}_{\ell-p} \rho_{\ell}}
\partial^{_{(v)}}_{\rho_{\ell}} -} \\[0.5cm]
{\displaystyle \sum^{2j}_{d=p+1} \Bigr (
\overline{\mu}^{\dot{\gamma}} \frac{\partial}{\partial \overline{\lambda}^{\dot{\gamma}}}
\bigl(A^{-1\dagger}_{(k)}\; \tilde{\sigma}_0\bigr)^{\dot{\beta}_{d-p} \; \rho_d}  \Bigr ) \,
\partial^{_{(v)}}_{\rho_d}
\prod^{p}_{n=1} (A_{(k)})^{\;\;\; \rho_n}_{\alpha_n}  \partial^{_{(v)}}_{\rho_n}
\prod^{2j}_{\underset{(\ell \neq d)}{\ell=p+1}}
 \bigl(A^{-1\dagger}_{(k)}\; \tilde{\sigma}_0\bigr)^{\dot{\beta}_{\ell-p} \rho_{\ell}}
 \partial^{_{(v)}}_{\rho_{\ell}}
\biggr)} = \\[0.5cm]
{\displaystyle \Bigl(  \lambda_{\gamma} \frac{\partial}{ \partial \mu_{\gamma}} - \overline{\mu}^{\dot{\gamma}} \frac{\partial}{\partial \overline{\lambda}^{\dot{\gamma}}} \Bigr)
\prod^p_{n=1}
(A_{(k)})^{\;\;\; \rho_n}_{\alpha_n}\, \partial^{_{(v)}}_{\rho_n}
\prod^{2j}_{\ell=p+1}
\bigl(A^{-1\dagger}_{(k)}\; \tilde{\sigma}_0\bigr)
^{\dot{\beta}_{\ell-p} \rho_{\ell}} }  \partial^{_{(v)}}_{\rho_{\ell}},
\end{array}
\ee
where to obtain equality we used the product rule. 
Further, substituting the right-hand side (\ref{Cheq6}) in (\ref{Cheq3}),
we are convinced of the validity of the relation (\ref{Smr3}).
\hfill \qed

{\bf Remark 1.}
Formulas from Proposition {\bf \ref{Smr}} can be used to construct tensors of arbitrary polarization
$m$, expressed in terms of Weyl spinors $\mu, \lambda$.
We first construct the polarization tensor $\overset{_{(m)}}{e}$ for $m=j$,
using the parametrization of the Wigner operators (\ref{npkp}) in terms of parameterization (\ref{aonC}) of operators
$A_{(k)}, A^{\dagger -1}_{(k)}$
\be \label{peVp}
\overset{_{(j)}}{e}^{(\dot{\beta_1}...\dot{\beta_r})}_{(\alpha_1...\alpha_p)}
 =\frac{(-1)^r}{(z)^p (z^*)^r} \mu_{\alpha_1} \cdots \mu_{\alpha_p} \overline{\lambda}^{\dot{\beta_1}} \cdots  \overline{\lambda}^{\dot{\beta_r}}.
\ee
Now, using recurrence formula (\ref{Smr3}) we can write the polarization tensor for $m=j-1$
\be \label{SMr31}
\overset{_{(j-1)}}{e}^{(\dot{\beta_1}...\dot{\beta_r})}_{(\alpha_1...\alpha_p)} =
\frac{1}{\sqrt{2j}} \frac{(-1)^r}{(z)^p (z^*)^r}
\bigl (\lambda_\gamma \frac{\partial}{\partial \mu_\gamma} -
\overline{\mu}^{\dot{\gamma}} \frac{\partial}{\partial \overline{\lambda}^{\dot{\gamma}}}\bigr)
\mu_{\alpha_1} \cdots \mu_{\alpha_p} \overline{\lambda}^{\dot{\beta_1}} \cdots  \overline{\lambda}^{\dot{\beta_r}}.
\ee
Further, applying the formula (\ref{Smr3}), one can obtain all the polarization tensors.

\section{Behrends-Fronsdal operator for $D$-dimensional case.\label{PrBeFr}}
\setcounter{equation}{0}

{\bf Defenition 1} { \it The Behrends-Fronsdal projection operator $\Theta(k)$
uniquely determined by the following conditions \newline

1) projective property and reality: $\;\;\;\;\;\Theta^2=\Theta,\;\; \Theta^\dagger=\Theta$; \newline

2) symmetry:	\;\;\;\;\;$\Theta^{n_1\cdots n_j}_{\cdots r_i \cdots r_\ell \cdots}=\Theta^{n_1\cdots n_j}_{\cdots r_\ell \cdots r_i \cdots},\;\;
\Theta^{\cdots n_i\cdots n_\ell \cdots}_{r_1 \cdots r_j}=\Theta^{\cdots n_\ell \cdots n_i\cdots}_{r_1 \cdots r_j}$; \newline

3) transversality:\;\;\;\;\;$k^{r_1}\Theta^{n_1 \cdots n_j}_{r_1 \cdots r_j}=0$,
$\;\;k_{n_1}\Theta^{n_1 \cdots n_j}_{r_1 \cdots r_j}=0$; \newline

4) tracelessness:\;\;\;\;\; $\eta^{r_1 r_2}\Theta^{n_1 \cdots n_j}_{r_1 r_2\cdots r_j}=0$. \newline
}

For the four-dimensional space-time $D=4$, the Behrends-Fronsdal projection operator $\Theta(k)$
for any spin $j$ was explicitly constructed in \cite{Fronsd}, \cite{BF}.
In \cite{IP}, \cite{PoTs}, \cite{FMoS} the generalization of the Behrends-Fronsdal operator to the case of an
arbitrary number of dimensions $D>2$ was found.
The construction was based on the properties of this operator,
which are listed in Defenition {\bf 1}.

Instead of the tensor $\Theta^{n_1 \dots n_j}_{r_1 \dots r_j}(k)$
symmetrized in the upper and lower indices,
was considered the generating function
 \be
 \lb{genT01}
 \Theta^{(j)}(x,y) = x^{r_1} \cdots x^{r_j} \,
 \Theta^{n_1 \dots n_j}_{r_1 \dots r_j}(k) \,
 y_{n_1} \cdots y_{n_j} \; .
 \ee
For concreteness, we assume that the tensor $\Theta(k)$ with components
 $\Theta^{n_1 \dots n_j}_{r_1 \dots r_j}(k)$ is defined in the pseudo-Euclidean $D$-dimensional
 space  $\mathbb{R}^{s,t}$
 $(s+t=D)$ with an arbitrary metric
 $\eta = ||\eta_{mn}||$,
 having the signature $(s,t)$. Indices
 $n_\ell$ and $r_\ell$ in (\ref{genT01})
 run through values $0,1,\dots,D-1$ and
 $(x_0,...,x_{D-1})$,
 $(y_0,...,y_{D-1}) \in \mathbb{R}^{s,t}$.

\begin{proposition}\label{svop1} (See  \cite{IP}) The generating function (\ref{genT01})
of the covariant projection operator
 $\Theta^{n_1 \dots n_j}_{r_1 \dots r_j}$
 (in $D$-dimensional space-time),
satisfying properties 1)-4), in Defenition {\bf 1}, has the form
 \be
 \lb{genT02}
 \Theta^{(j)}(x,y) = \sum_{A=0}^{[\frac{j}{2}]} a^{(j)}_A \;
 \bigl(\Theta^{(y)}_{(y)} \, \Theta^{(x)}_{(x)} \bigr)^A  \;
 \bigl(\Theta^{(y)}_{(x)} \bigr)^{j -2A}  \; ,
 \ee
 where $[\frac{j}{2}]$ -- integer part of $j/2$,
the coefficients $a^{(j)}_{A}$
satisfy recurrent relation
 \be
 \lb{genT10lk}
 a^{(j)}_{A} = - \frac{1}{2} \; \frac{(j -2A+2)(j -2A+1)}{
 A \; (2j -2A+D-3)} \; a^{(j)}_{A-1} \; .
 \ee
The solution of equation (\ref{genT10lk}) has the form
  \be
 \lb{genT11}
 a^{(j)}_{A} = \Bigl( - \frac{1}{2}\Bigr)^A \frac{j!}{
 (j -2A)! \, A! \, (2j +D-5)(2j +D-7)\cdots (2j +D -2A -3)}  \; ,
 \;\;\; (A \geq 1) \; ,
 \ee
 $a^{(j)}_{0}=1$, 
and the function
 $\Theta^{(y)}_{(x)}$ is defined as follows
 ($\eta_{r n}$ -- the metric of space $\mathbb{R}^{s,t}$):
 \be
 \lb{genT}
 \Theta^{(y)}_{(x)} \equiv  \Theta^{(1)}(x,y) =  x^r \, y_n \, \Theta^{n}_{r} \; ,
 \;\;\;\; \Theta_{r}^{n}=\eta_{r}^{n}-\frac{k_r k^n}{k^2} \; .
 \ee
\end{proposition}
Now we will prove some new statement about the generation fuction of the operator $\Theta^{(j)}$.  
\begin{proposition} \label{rfh0}
For the generation function (\ref{genT01}) the following recurrence formula is hold
\be \lb{shreq3}
\begin{array}{c}
\Theta^{(j)}(x,y) = \frac{1}{j!} x^{n_1} \cdots x^{n_j} \,
(\overline{\amalg \!\! \amalg}_{j-1})^{\ell_1 \dots \ell_{j-1} \ell_j}_{n_1 \dots n_{j-1} n_j} \,
\frac{\partial}{\partial z^{\ell_1}} \cdots \frac {\partial} {\partial z^{\ell_{j-1}}} \, \frac{\partial}{\partial t^{\ell_{j}} }\, \Theta^{(j-1)}(z,y) \, \Theta(t, y).
\end{array}
\ee
Here we defined element
\be \lb{newsh}
\overline{\amalg \!\! \amalg}_{j-1}  = 
\bigl (\tilde{1}+\tilde{\tau}_{j-1} + \tilde{\tau}_{j-2} \tilde{\tau}_{j-1} + \dots + \tilde{\tau}_{j-k} \cdots \tilde{\tau}_{j-2} \tilde{\tau}_{j-1} + \dots
+\tilde{\tau}_1 \tilde{\tau}_2 \cdots \tilde{\tau}_{j-1} \bigr) \, .
\ee
where $\tilde{1} = \Theta_{n_1}^{l_1} \cdots \Theta_{n_j}^{l_j} $ and element $\tilde{\tau}_{i}$ have the form
\footnote{Here the operator $\tilde{1}$ really plays the role of a unit, since it trivially 
acts on all covariant combinations constructed from $\Theta=||\Theta^{\,n}_{m}||$ on (\ref{genT}) and space-time metric $\eta = ||\eta_{nm}||$ .}
\be \lb{dtau1}
\begin{array}{c}
(\, \tilde{\tau}_i \,)_{n_1 \dots n_j}^{l_1 \dots l_j} = \Bigl ( \Theta_{n_1}^{l_1} \cdots  \Theta_{n_{i+1}}^{l_i}  \Theta_{n_i}^{l_{i+1}} \cdots \Theta_{n_j}^{l_j} 
- \frac{2}{ (\omega + 2(i - 1) )} \cdot
\Theta_{n_1}^{l_1} \cdots  \Theta_{n_i n_{i+1}}  \Theta^{l_i l_{i+1}} \cdots \Theta_{n_j}^{l_j} 
\Bigr ),
\end{array}
\ee
where  $\omega = D-1$. 
\end{proposition}
{\bf Proof.}
First we simplify the formula (\ref{shreq3}).
Note, we will need it later, that the following formula is hold.
\be \lb{rqsh}
\overline{\amalg \!\! \amalg}_{j-1}  = 
\bigl(\tilde{1}+\overline{\amalg \!\! \amalg}_{j-2} \, \tilde{\tau}_{j-1} \bigr) \, .
\ee
Consider the differential operator
\be \lb{dish}
x^{n_1} \cdots x^{n_j} \,
(\overline{\amalg \!\! \amalg}_{j-1})^{\ell_1 \dots \ell_{j-1} \ell_j}_{n_1 \dots n_{j-1} n_j} \,
\frac{\partial}{\partial z^{\ell_1}} \cdots \frac {\partial} {\partial z^{\ell_{j-1}}} \, \frac{\partial}{\partial t^{\ell_{j}} }
\ee 
from the right hand-side (\ref{shreq3}).
Next we show,
that (\ref{dish}) reduces to the following sum of two simple terms
\be \lb{dish1} 
\begin{array}{c}
j \cdot \Theta^{(\,x\,)}_{(\partial_t)} \, \bigr ( \Theta^{(\,x\,)}_{(\partial_z)} \bigl )^{j-1} - \,\, \frac{j(j-1)}{(\omega + 2 (j-2))} \, \Theta^{(x)}_{(x)} \, 
\Theta^{(\partial_t)}_{(\partial_z)} \, \bigl ( \Theta^{(\,x\,)}_{(\partial_z)} \bigr)^{j-2}
\end{array}
\ee 
Then using (\ref{dish1}) we can rewrite the formula (\ref{shreq3})
\be \lb{shreq4}
\begin{array}{c}
\Theta^{(j)}(x,y) = \frac{1}{(j-1)!} 
\Bigl ( \Theta^{(x)}_{(y)} \, \bigr ( \Theta^{(\,x\,)}_{(\partial_z)} \bigl )^{j-1} 
- \,\, \frac{(j-1)}{(\omega + 2 (j-2))} \, \Theta^{(x)}_{(x)} \, 
\Theta^{(\,y\,)}_{(\partial_z)} \, 
\bigl ( \Theta^{(\,x\,)}_{(\partial_z)} \bigr)^{j-2} \Bigr)
\, \Theta^{(j-1)}(z,y)\,,
\end{array}
\ee
here we immediately differentiated by the variable $t$. 

We now show that the formula (\ref{dish1}) is equivalent to (\ref{dish}). 
We carry out the proof by induction on $j$. For $j=2$ the formula (\ref{newsh}) is represented as
\be \lb{in1}
(\overline{\amalg \!\! \amalg}_{1})^{r_1 r_2}_{n_1 n_2} = \Theta_{n_1}^{r_1} \Theta_{n_2}^{r_2} +  \Theta_{n_2}^{r_1} \Theta_{n_1}^{r_2} 
- \frac{2}{\omega}\, \Theta_{n_1 n_2} \Theta^{r_1 r_2}
\ee
We make a contraction similar to (\ref{dish}) for $j=2$, using the formula (\ref{in1})
and the definition of the generating function $\Theta^{(x)}_{(y)}$ из (\ref{genT}), as a result we have
\be \lb{in2}
x^{n_1} x^{n_2} (\overline{\amalg \!\! \amalg}_{1})^{r_1 r_2}_{n_1 n_2} \frac{\partial}{\partial z^{r_1}} \frac{\partial}{\partial t^{r_2}} =
2 \cdot \Theta^{(\,x\,)}_{(\partial_t)} \, \Theta^{(\,x\,)}_{(\partial_z)}-  \frac{2}{\omega} \, \Theta^{(x)}_{(x)} \, \Theta^{(\partial_t)}_{(\partial_z)} ,
\ee
it follows that the formulas  (\ref{dish1}) and (\ref{dish}) is equivalent for $j=2$. 
 Now we consider the sequence of the transformations for the (\ref{dish}) 
in case any $j$
\be \lb{in3}
\begin{array}{c}
x^{n_1} \cdots x^{n_j} \,
(\overline{\amalg \!\! \amalg}_{j-1})^{\ell_1 \dots \ell_{j-1} \ell_j}_{n_1 \dots n_{j-1} n_j} \,
\frac{\partial}{\partial z^{\ell_1}} \cdots \frac {\partial} {\partial z^{\ell_{j-1}}} \, \frac{\partial}{\partial t^{\ell_{j}} } =
x^{n_1} \cdots x^{n_j} \, \bigl ( \Theta_{n_1}^{\ell_1} \cdots \Theta_{n_j}^{\ell_j} + \\ [0.5cm]
(\overline{\amalg \!\! \amalg}_{j-2})^{\ell_1 \dots \ell_{j-2} r_{j-1}}_{n_1 \dots n_{j-2} n_{j-1}} \Theta_{n_j}^{r_j} ( 
  \Theta_{r_j}^{\ell_{j-1}} \Theta_{r_{j-1}}^{\ell_j} 
- \frac{2}{(\omega +2(j-2) )}\,  \Theta_{r_{j-1} r_j} \Theta^{\ell_{j-1} \ell_j} ) \bigr)
\frac{\partial}{\partial z^{\ell_1}} \cdots \frac {\partial} {\partial z^{\ell_{j-1}}} \, \frac{\partial}{\partial t^{\ell_{j}} } = \\[0.5cm]
=\Theta^{(\,x\,)}_{(\partial_t)} \, \bigr ( \Theta^{(\,x\,)}_{(\partial_z)} \bigl )^{j-1} 
\, + \, (j-1) \Bigl( \Theta^{(\,x\,)}_{(\partial_c)} \, \bigr ( \Theta^{(\,x\,)}_{(\partial_z)} \bigl )^{j-2} - \,\, \frac{(j-2)}{(\omega + 2 (j-3))} \, \Theta^{(x)}_{(x)} \, 
\Theta^{(\partial_c)}_{(\partial_z)} \, \bigl ( \Theta^{(\,x\,)}_{(\partial_z)} \bigr)^{j-3} \Bigr )  \\[0.5cm]
\cdot \Bigl ( 
  \Theta^{(\,x\,)}_{(\partial_z)} \Theta_{(\,c\,)}^{(\partial_t)} 
- \frac{2}{(\omega +2(j-2) )}\,  \Theta_{(c)}^{(x)} \Theta^{(\partial_t)}_{(\partial_z)}) \Bigr) = \Theta^{(\,x\,)}_{(\partial_t)} \, \bigr ( \Theta^{(\,x\,)}_{(\partial_z)} \bigl )^{j-1} 
+(j-1)  \Bigl (  \Theta^{(\,x\,)}_{(\partial_t)}\, \bigr ( \Theta^{(\,x\,)}_{(\partial_z)} \bigl )^{j-1} -  \\[0.5cm]
- \bigl ( \frac{2}{(\omega +2(j-2) )}\,
+ \,\frac{(j-2)}{(\omega + 2 (j-3))} - \frac{2 (j-2)} {(\omega +2(j-2) ) (\omega + 2 (j-3))} \bigr) \, \Theta^{(x)}_{(x)} \, 
\Theta^{(\partial_t)}_{(\partial_z)} \, \bigl ( \Theta^{(\,x\,)}_{(\partial_z)} \bigr)^{j-2} \Bigr )  = \\[0.5cm]
= j \cdot \Theta^{(\,x\,)}_{(\partial_t)} \, \bigr ( \Theta^{(\,x\,)}_{(\partial_z)} \bigl )^{j-1} - \,\, \frac{j(j-1)}{(\omega + 2 (j-2))} \, \Theta^{(x)}_{(x)} \, 
\Theta^{(\partial_t)}_{(\partial_z)} \, \bigl ( \Theta^{(\,x\,)}_{(\partial_z)} \bigr)^{j-2} \, ,
\end{array}
\ee 
here in the first equality we used the recurrence relation (\ref{rqsh}), 
in the second equality we used the induction hypothesis  (equivalence of formulas  (\ref{dish1}) and (\ref{dish}) for $\overline{\amalg \!\! \amalg}_{j-2}$), 
in the third equation we applied differentiation with respect to the auxiliary $D$-vector $c$.

We now show that the formula (\ref{shreq4}) is valid for the generating function (\ref{genT02}). 
We act with the differential operator from the right-hand side (\ref{shreq4}) on the generating function $\Theta^{(j-1)}(z,y)$,
taken as a series (\ref{genT02}) 
\be \lb {chs0}
\begin{array}{c}
\Theta^{(j-1)}(z,y) = 
\displaystyle{ \sum_{A=0}^{[\frac{j-1}{2}]} a_A^{(j-1)}  
\bigl(\Theta^{(y)}_{(y)} \, \Theta^{(z)}_{(z)} \bigr)^A  \;
 \bigl(\Theta^{(y)}_{(z)} \bigr)^{j-1-2A}}  \; ,
\end{array}
\ee
were the coefficients $a_A^{(j-1)}$ define by formula (\ref{genT11}). 
\newline
The result of the action will be the following components
\be \lb{shs1}
\begin{array}{c}
\displaystyle{\sum_{A=0}^{[\frac{j-1}{2}]} a_A^{(j-1)}  
\bigl(\Theta^{(y)}_{(y)} \, \Theta^{(x)}_{(x)} \bigr)^A  \;
 \bigl(\Theta^{(y)}_{(x)} \bigr)^{j-2A}} \\[0.5cm]

\frac{1}{(\omega + 2(j-2))} \cdot \displaystyle{\sum_{A=0}^{[\frac{j-1}{2}]} 2A \cdot a_A^{(j-1)}  
\bigl(\Theta^{(y)}_{(y)} \, \Theta^{(x)}_{(x)} \bigr)^A  \;
 \bigl(\Theta^{(y)}_{(x)} \bigr)^{j-2A}} \\[0.5cm]

\frac{1}{(\omega + 2(j-2))} \cdot \displaystyle{\sum_{A=0}^{[\frac{j-1}{2}]} (j-1-2A) \cdot a_A^{(j-1)}  
\bigl(\Theta^{(y)}_{(y)} \, \Theta^{(x)}_{(x)} \bigr)^{A+1}  \;
 \bigl(\Theta^{(y)}_{(x)} \bigr)^{j-2(A+1)}}
\end{array}
\ee
In the proof, we will consider the case when $j-1$ is an even number, 
the proof for odd $j-1$  is carried out in a similar way.
Let's make in each line (\ref{shs1}) some transformations
\be \lb{shs3}
\begin{array}{c} 
 a_0 \, \bigl(\Theta^{(y)}_{(x)} \bigr)^{j} + \displaystyle{\sum_{A=1}^{[\frac{j}{2}]} a_A^{(j-1)}  
\bigl(\Theta^{(y)}_{(y)} \, \Theta^{(x)}_{(x)} \bigr)^A  \;
 \bigl(\Theta^{(y)}_{(x)} \bigr)^{j-2A}} \\[0.5cm]

-\frac{1}{(\omega + 2(j-2))} \cdot \displaystyle{\sum_{A=1}^{[\frac{j}{2}]} 2A \cdot a_A^{(j-1)}  
\bigl(\Theta^{(y)}_{(y)} \, \Theta^{(x)}_{(x)} \bigr)^A  \;
 \bigl(\Theta^{(y)}_{(x)} \bigr)^{j-2A}} \\[0.5cm]

-\frac{1}{(\omega + 2(j-2))} \cdot \displaystyle{\sum_{A=1}^{[\frac{j}{2}]} (j+1-2A) \cdot a_{A-1}^{(j-1)}  
\bigl(\Theta^{(y)}_{(y)} \, \Theta^{(x)}_{(x)} \bigr)^{A}  \;
 \bigl(\Theta^{(y)}_{(x)} \bigr)^{j-2A}}
\end{array}
\ee
In the first line we excluded the first term from the total sum 
and we took into account the fact that
for even $j-1$ the relation $[\frac{j-1}{2}] = [\frac{j}{2}]$ is hold.
In the second line, we again used the equality $[\frac{j-1}{2}] = [\frac{j}{2}]$
and eliminated the obviously zero first term.
In the third line, we eliminated the last term, then made a shift of the summation parameter $A \rightarrow A-1$,
and used  $[\frac{j-1}{2}] = [\frac{j}{2}]$. Now we add all terms (\ref{shs3})
as a result we get 
\be \lb{shs4}
a_0^{(j-1)} \, \bigl(\Theta^{(y)}_{(x)} \bigr)^{j} +  \displaystyle{\sum_{A=1}^{[\frac{j}{2}]} B_A 
\bigl(\Theta^{(y)}_{(y)} \, \Theta^{(x)}_{(x)} \bigr)^A  \;
 \bigl(\Theta^{(y)}_{(x)} \bigr)^{j-2A}}
\ee
where the coefficient $B_A$ is determined by the following chain of equalities
\be \lb {shs5}
\begin{array}{c}
B_A = a_A^{(j-1)} - \frac{2A}{(\omega + 2(j-2))} a_A^{(j-1)} - \frac{(j-2A+1)}{(\omega + 2(j-2))} a_{A-1}^{(j-1)} = \\[0.5cm]
=  a_A^{(j-1)} \Bigl ( \frac{ \omega + 2(j-2) - 2A}{(\omega + 2(j-2) )} \Bigl ) - \frac{(j-2A + 1)}{(\omega + 2(j-2) )}  a_{A-1}^{(j-1)}  = \\[0.5cm]
= -\frac{1}{(\omega +2 (j-2) )} \Bigl ( \frac{1}{2} \frac{(j-2A+1)(j-2A)}{A} + (j-2A+1) \Bigr ) a_{A-1}^{(j-1)} 
= -\frac{1}{2} \frac{j (j-2A +1)}{A (\omega +2 (j-2) )} a_{A-1}^{(j-1)},
\end{array}
\ee
here we used the recurrence relation (\ref{genT10lk}) for the coefficients $a^{(j-1)}_{A}$. 
Substituting now the explicit expression for $a_{A-1}^{(j-1)}$, we see that $B_A$ exactly coincides with $a^{(j)}_A$. 
As a result, we can write
\be \lb{shs6}
\Theta^{(j)}(x,y) = \displaystyle{\sum_{A=0}^{[\frac{j}{2}]} a_A^{(j)}  
\bigl(\Theta^{(y)}_{(y)} \, \Theta^{(x)}_{(x)} \bigr)^A  \;
 \bigl(\Theta^{(y)}_{(x)} \bigr)^{j-2A}}
\ee
\hfill \qed

\section{Conclusion}
\setcounter{equation}0

We hope that the formalism considered in this paper for describing massive particles
of arbitrary spin will be useful in the construction of scattering amplitudes of massive particles
in a similar way to the construction of spinor-helicity scattering amplitudes
for massless particles \cite{Witt},  \cite{ElHua}.
Some steps in this direction
have already been done in papers \cite{CoMa},\cite{CJM},\cite{Ma} where
the analogous formalism and its special generalization were used. 
We also think that using the methods from \cite {IM} the formulas (\ref{shreq3}) - (\ref{dtau1}) can be
 generalized for projection operators of any type of symmetry (corresponding to arbitrary Young diagrams).

The first part of this work was supported by RFBR, grant 18-52-05002 Арм\_а; the second part was supported by RFBR, grant 19-01-00726А. 

The author is grateful to A.P. Isaev for formulation of the problem and helpful discussions. 

 \vspace{1cm}

\begin{otherlanguage}{english}

\end{otherlanguage}
\end{document}